\documentclass[conference]{IEEEtran}
\IEEEoverridecommandlockouts
\usepackage{comment}
\usepackage{cite}
\usepackage{amsmath,amssymb,amsfonts}
\usepackage{algorithmic}
\usepackage{graphicx}
\usepackage{textcomp}
\usepackage{xcolor}
\usepackage{url}
\usepackage{soul}
\usepackage{todonotes}
\def\BibTeX{{\rm B\kern-.05em{\sc i\kern-.025em b}\kern-.08em
    T\kern-.1667em\lower.7ex\hbox{E}\kern-.125emX}}

\begin{document}

\title{Bug Whispering: Towards Audio Bug Reporting}

\author{\IEEEauthorblockN{Elena Masserini}
\IEEEauthorblockA{\textit{University of Milano-Bicocca}\\
Milan, Italy \\
elena.masserini@unimib.it}
\and
\IEEEauthorblockN{Daniela Micucci}
\IEEEauthorblockA{\textit{University of Milano-Bicocca}\\
Milan, Italy \\
daniela.micucci@unimib.it}
\and
\IEEEauthorblockN{Leonardo Mariani}
\IEEEauthorblockA{\textit{University of Milano-Bicocca}\\
Milan, Italy \\\
leonardo.mariani@unimib.it}
}

\maketitle

\begin{abstract}
Bug reporting is a key feature of mobile applications, as it enables developers to collect information about faults that escaped testing and thus affected end-users. This paper explores the idea of allowing end-users to immediately report the problems that they experience by recording and submitting audio messages. Audio recording is simple to implement and has the potential to increase the number of bug reports that development teams can gather, thus potentially improving the rate at which bugs are identified and fixed. However, audio bug reports exhibit specific characteristics that challenge existing techniques for reproducing bugs. This paper discusses these challenges based on a preliminary experiment, and motivates further research on the collection and analysis of audio-based bug reports.
\end{abstract}


\section{Introduction}

Effective failure reporting is important since it enables users to promptly notify the failures experienced in the field to the development team, and allows developers to collect information about the bugs that escaped testing, 
simplifying the progressive release of 
improved versions of their software~\cite{Zimmermann:GoodBugReport:TSE:2010}. 

The most common means of failure reporting are \textit{crash reporting systems} and \textit{issue tracking systems}. Crash reporting systems are automated tools that detect unhandled exceptions and crashes at runtime and send reports with diagnostic data. Popular crash reporting systems are Sentry\footnote{\url{https://sentry.io/welcome/}}, Firebase Crashlytics\footnote{\url{https://firebase.google.com/docs/crashlytics}}, and BugSnag\footnote{\url{https://www.bugsnag.com/}}, to developers. Issue tracking systems, such as, GitHub Issues\footnote{\url{https://github.com/features/issues}}, Google Issue Tracker\footnote{\url{https://issuetracker.google.com/issues}}, and Bugzilla\footnote{\url{https://www.bugzilla.org/}}, are reporting platforms where users can submit reports about the bugs they encountered, describing the steps to reproduce the failures, sometimes enriched with screenshots and videos.


Crash reporting systems are indeed very practical, since they are mostly automated (users just have to authorize the uploading of the report). However, they are \emph{limited both in usefulness}, since reports tend to include limited information about the failures (e.g., only the program stacktrace with a snapshot of the memory), and in \emph{scope}, since they cannot be applied to non-crashing failures (e.g., cases where an application simply outputs an incorrect result).

On the other hand, issue tracking systems support detailed reporting of all types of failures, with a range of automated failure reproduction techniques for bug reports that have been studied so far, such as ReCDroid\cite{recdroid}, Yakusu\cite{yakusu} and ScopeDroid\cite{scopedroid}. However, issue tracking systems \emph{require effort and a certain level of expertise} to be used, discouraging the less experienced or less motivated users. Indeed, the average app user does not report failures on an issue tracking system.

An appealing bug reporting option that has not been thoroughly investigated yet is \emph{audio-based bug reporting}, where users can directly record voice messages about an experienced issue, as soon as it occurs. This reporting mode effectively combines \emph{practicality}, since it is based on the recording of an audio that is a very natural, non-technical, and well-known interaction modality, with \emph{completeness}, since users can potentially provide rich descriptions about the steps that caused a failure, and \emph{broad scope of application}, since it can be applied to any type of issue, including non-crashing failures.   

Developers would receive several complementary descriptions of the same problems, quickly obtaining an accurate understanding of the bugs that need to be fixed. Compared to text, audio also discloses information about the emotional status of the users (e.g., the sentiment of the user at the time the failure was reported), which can be used to infer information about the user satisfaction and the impact of bugs.

Collecting audio bug reports does not introduce significant technical challenges, since this feature can be easily added to apps. The main research challenges lie in two aspects: (1) understanding how the content of audio bug reports compares to traditional bug reports, and (2) defining techniques to analyze audio reports, to support bug reproduction and fixing.

This fast abstract addresses the first challenge  
by using an audio bug reporting module that we implemented to collect preliminary evidence about the style and content of audio reports. Our preliminary results reveal interesting tradeoffs between audio reports and textual issues, and highlight the need for further research on audio bug reporting.

\section{Audio Bug Reporting}

An audio bug reporting approach allows users to submit audio-based reports of issues experienced in the field. The main components of an audio report module are an \emph{audio recording module}, which can be activated by users through a specific action (e.g., a voice command, a dedicated gesture, a button click) as soon as a problem occurs, and a \emph{composite reporting module}, which creates a bundle with the recorded audio, a screenshot of the problematic app, and metadata collected from the device (e.g., the device model, the version of the operating system, and the version of the app). 

\begin{figure}[h]
    \centering
    \includegraphics[width=0.8\linewidth]{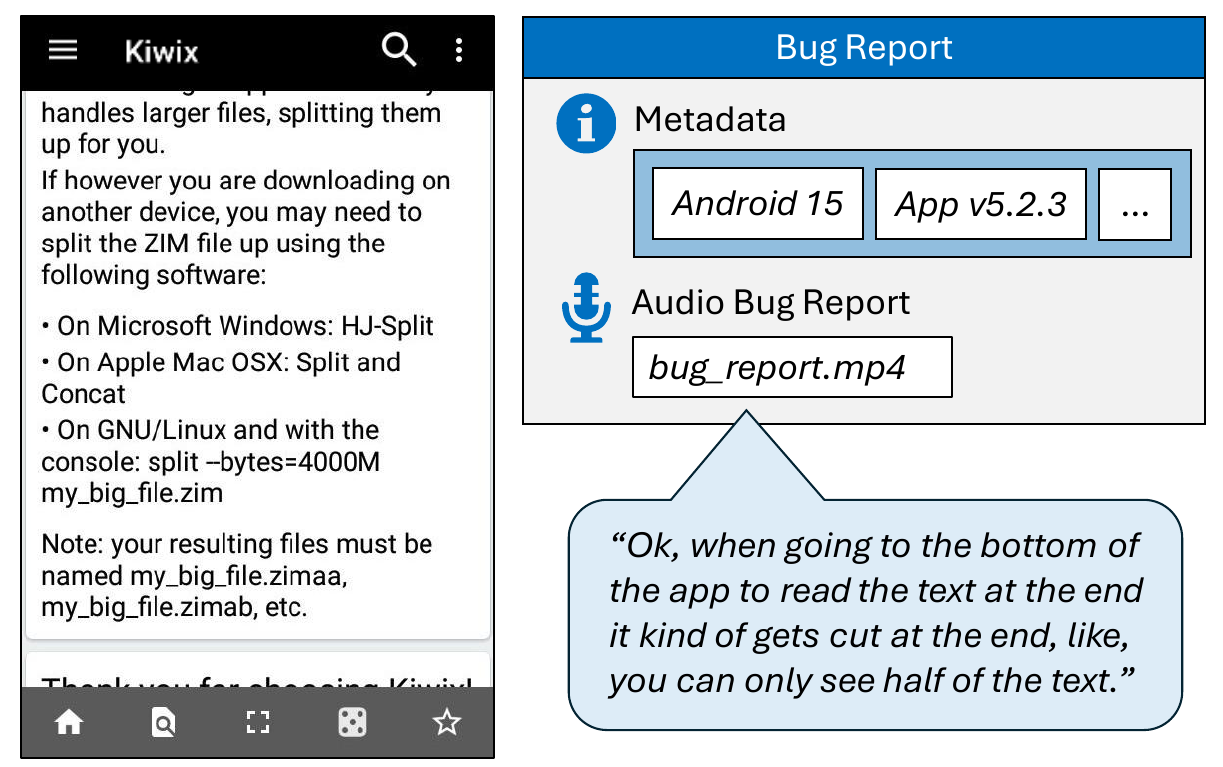}
    \caption{Example audio bug report in the Kiwix application}
    \label{fig:audio-bug-report}
\end{figure}

We implemented a prototype version of this module that can be injected directly into existing Android apps using Frida\footnote{\url{https://frida.re/docs/android/}}. The user can activate audio recording through a triple tap on the screen. Figure \ref{fig:audio-bug-report} shows an example audio bug report submitted by a participant in our preliminary study describing a visualization bug in Kiwix\footnote{\url{https://github.com/kiwix/kiwix-android/issues/555}}.

\section{Preliminary Empirical Evidence}

To understand the specific traits in content and style that may distinguish audio bug reports from traditional bug reports, we designed and executed a preliminary small-scale study with ten mobile app users randomly selected from the students of the University of Milano - Bicocca. We selected ten GitHub issues reporting non-crashing failures from the \textsc{AndroR2} dataset~\cite{Wendland_2021}, and we installed and configured the faulty version of each app on an Android device. Participants were asked to interact with these applications by performing the steps required to reproduce the failure, without being given information about the expected malfunction. When the bug occurred, the participants submitted an audio report using our bug reporting module. Each participant engaged with five applications, for an overall total of 50 bug reports collected (five reports per bug). The audio recordings were later manually transcribed and compared to the corresponding issues available on GitHub. 
We observed missing content, extra content, and stylistic differences, which are discussed below. 

 \textbf{Content missing in audio bug reports compared to issues}\newline 
\emph{Missing steps necessary to reproduce the failure}. Less than half of the audio reports describe a sequence of two or more steps to reproduce failures, while the rest of the reports only report the last action executed before the failure (e.g., ``\emph{When I take a photo it doesn't appear in the event icon}") or, less common, no step at all (e.g., ``\emph{The text went into all caps with no reason, I didn't mean to do that}"). Although online issues may also contain limited information about the failure-reproducing steps~\cite{Huang-2025}, audio bug reports exacerbate this problem.

\emph{Missing details about steps.}
Audio bug reports tend to describe interactions in a less precise way (e.g., ``\emph{The button that changes characters}") than issues, where views are often referred with precise indications about text and position (e.g., ``\emph{The symbol layout button labeled as 123?}"\footnote{\url{https://github.com/Helium314/HeliBoard/issues/1256}})

\textbf{Extra content in audio bug reports compared to issues}\newline
\emph{User sentiment is recognizable from audio report}. Audio reports reveal more insights about the users' feelings and experiences, in a way that written reports, usually more formal and standardized, fail to capture. 
A collection of audio reports can thus provide developers with valuable insights into the urgency of issues as well as the overall satisfaction of users.

\textbf{Stylistic differences}\newline
\emph{Lack of structure}. Issues tend to be longer and more structured, sometimes even based on a given template, and therefore more complete and explicit. Audio bug reports instead are spontaneous and unconstrained narratives, without structure, where important information is often implicit. 

\emph{Implicit expected behavior}. The expected behavior is often left implicit in bug reports.
For example, the report ``\emph{When rotating the screen I see many logs but I don't understand}" requires making deductions to identify the expected behaviour, in comparison to the correspondingg issue description that explicitly refers to
``\emph{No new logging file on landscape change}"\footnote{\url{ https://github.com/barbeau/gpstest/issues/404}}. 

\emph{Colloquial and imprecise language} The language in audio bug reports is generally more informal than issues, and include colloquial terms and periphrasis (e.g., ``\emph{I am kind of stuck}"), and disfluencies typical of spoken language like repetitions, interlayers, and incomplete sentences (e.g., ``\emph{...the text is so, is too, it so long that...}", ``\emph{...and, well, the title ..}").

\section{Conclusions}

In this work we introduced audio bug reporting and presented a preliminary study comparing audio bug reports to traditional issue reports.
Preliminary results show that audio bug reports have specific characteristics, in content and style, that call for more research in bug reporting and failure reproduction. 

\bibliographystyle{IEEEtran}
\bibliography{IEEEabrv,references}

\end{document}